\documentclass[inte,nonblindrev]{preprint} 

\OneAndAHalfSpacedXII 

\usepackage{natbib}
 \bibpunct[, ]{(}{)}{,}{a}{}{,}%
 \def\bibfont{\small}%

\usepackage{color}
\definecolor{EditBlue}{RGB}{0,0,0}
\newcommand{\eb}[1]{{\color{EditBlue}#1}}

\definecolor{EditGreen}{RGB}{0,0,0}
\newcommand{\eg}[1]{{\color{EditGreen}#1}}

\usepackage{multirow}
\usepackage[table,xcdraw]{xcolor}
\usepackage{booktabs}

\TheoremsNumberedThrough     

\EquationsNumberedThrough    

\begin{document}



\RUNTITLE{Estimating Effectiveness of Identifying Trafficking}

\TITLE{Estimating Effectiveness of Identifying Human Trafficking via 
Data Envelopment Analysis}

\ARTICLEAUTHORS{%
\AUTHOR{Geri L. Dimas}
\AFF{Data Science Program, Worcester Polytechnic Institute}
\AUTHOR{Malak El Khalkhali}
\AFF{Business School, Worcester Polytechnic Institute}
\AUTHOR{Alex Bender}
\AFF{Mechanical and Industrial Engineering Department, Northeastern University}
\AUTHOR{Kayse Lee Maass}
\AFF{Mechanical and Industrial Engineering Department, Northeastern University}
\AUTHOR{Renata A. Konrad}
\AFF{Business School, Worcester Polytechnic Institute}
\AUTHOR{Jeffrey S. Blom}
\AFF{Love Justice International}
\AUTHOR{Joe Zhu}
\AFF{Business School, Worcester Polytechnic Institute}
\AUTHOR{Andrew C. Trapp}
\AFF{Business School, Worcester Polytechnic Institute}
\AFF{Data Science Program, Worcester Polytechnic Institute}
} 

\ABSTRACT{%
Transit monitoring is a preventative approach used to identify possible cases of human trafficking \eg{prior to exploitation} while an individual is in transit or before one crosses a border. Transit monitoring is often conducted by non-governmental organizations (NGOs) who train staff to identify and intercept suspicious activity. Love Justice International (LJI) \eb{is a well-established} NGO that has been conducting transit monitoring for years along the Nepal-India border at \eg{multiple} monitoring stations. In partnership with LJI, we developed a system that uses data envelopment analysis (DEA) to help LJI decision-makers evaluate the performance of these stations \eg{at intercepting potential human-trafficking victims given the amount of resources (e.g. staff, etc.) available} and make specific operational improvement recommendations. \eb{ Our model consists of 91 decision-making units (DMUs) from 7 stations over 13 quarters and considers three inputs, four outputs, and 3 homogeneity criteria. Using this model w}e identified efficient stations, compared rankings of station performance, and recommended strategies to improve efficiency. To the best of our knowledge, this is the first application of DEA in the anti-human trafficking domain.
}%



\maketitle

%


Human trafficking involves the commercial exchange and exploitation of humans for monetary gain or benefit and constitutes a gross violation of human rights \citep{Gajic_Veljanoski_2007}. In 2016, \eb{an} estimated 24.9 million people globally were being trafficked for sex and/or labor \citep{ILO_2017}. Trafficking occurs both domestically and internationally in all regions of the world where individuals are made vulnerable by environments of poverty, conflict, natural disaster, unemployment, and desperation \citep{USDS_2019}. Victims range from child soldiers and child brides to domestic workers (e.g. housekeepers), forced laborers (in occupations including commercial fishing, manufacturing, construction, mining, and agricultural work, to name a few), people in the commercial sex industry, and beggars (\citealp{Feingold_2005}, \citealp{Logan_2009}). Forced labor and sexual exploitation generate an estimated \$150 billion (U.S.) globally in illegal profits each year \citep{ILO_2012}. In addition to economic impacts, trafficking has severe health implications. Physical and sexual violence are common among survivors of sex trafficking, as is profound psychological manipulation. Labor trafficking victims often suffer from exhaustion, dehydration, heat strokes, hypothermia, respiratory issues, and skin infections \citep{Napolitano_2016}. Trafficked survivors may experience complex trauma as a result of forced isolation, deprivation, psychological abuse, degradation, and threats made to themselves or others \citep{Greenbaum_2014}.

To address the pervasiveness of human trafficking, most countries have enacted laws and policies to prosecute those who traffic humans and to provide assistance and protection to survivors of human trafficking. Awareness of, and efforts to, combat human trafficking are increasing. While \eb{anti-}human trafficking efforts \eb{to date have largely} focused on the sex trafficking of women, recent efforts have acknowledged that trafficking is a complex system that affects people of all genders and ages in a wide range of industries \citep{Denton_2010}.

Despite the social, moral, and economic need to address human trafficking, resources to address this issue remain extremely limited \citep{Foot_2020}. While the gap between the current funding of anti-trafficking interventions and needs is challenging to quantify, the literature consistently refers to the inadequacy of current funding for human trafficking interventions (\citealt{Cockayne_2015}, \citealt{Moses_2016}, \citealt{Idris_2017}). A critical challenge for agencies and governments is to use scarce resources well, as well as to effectively evaluate the impact of their anti-trafficking initiatives. Even so, many anti-human trafficking interventions continue to operate without an adequate evidence base (\citealt{Foot_2015}, \citealt{Davy_2016}). Though the anti-trafficking community largely lacks the analytical expertise and resources necessary to evaluate the effective allocation of resources, there is a tremendous opportunity to i) increase quality evaluations of anti-human trafficking programs\eb{;} ii) ensure that programs are targeted, implemented, and delivered effectively, and iii) improve the knowledge concerning the impact of interventions \citep{Davy_2016}.

Data analytics \eb{can be used to} improve the lives of individuals (\citealt{Raghupathi_2014}, \citealt{Crowley_2017}), \eb{support} decision\eb{-making in practice} \citep{Mandinach_2012}, and prove the value of routine collection for operational decision making (\citealt{UNHCR_2013}, \citealt{Boswell_2016}). Despite evidence suggesting that the awareness of the value of routinely collected data may be increasing (\citealt{Okun_2013}, \citealt{Jorm_2015}), the benefits gained from data analysis in the nonprofit and humanitarian sectors are rarely documented. More alarmingly, strategies for data collection and management to support such analyses are borrowed from the for-profit sector and do not address the unique needs of humanitarian nonprofits, such as collecting data \eb{regarding} vulnerable populations in complex environments with limited funding\eb{. Furthermore, data in nonprofit and humanitarian sectors} are not conducive to clean or simple data collection. In the severely resource-constrained environments in which nonprofits \eb{typically} operate, data analyses to support operational strategy and decision making can have positive effects such as improving finances, increasing operational efficiency, and enhancing donor relationships. Though many donors “require an assessment of the deployment and performance improvements resulting from their investments,” and continuous improvement, it can be difficult to develop the processes, structures, and systems necessary to support strategy and decision making as nonprofits typically do not have the means \citep{Berenguer_2016}. Routinely collected and analyzed data present a valuable and underutilized opportunity for nonprofits.

Performance assessment and monitoring is critical to establishing benchmarks for best practices and guiding improvement recommendations. Data envelopment analysis (DEA) is a \eb{data-driven} analytical tool used to assess the performance of units within an organization, or across organizations. Such analytics can highlight possible improvements in effectiveness for organizations, including those involved in anti-trafficking operations. We use DEA to \eb{analyze existing data and} evaluate the performance of border stations of non-governmental organization (NGO) Love Justice International (LJI) in Nepal engaged in the trafficking intervention strategy known as transit-monitoring. \eg{The performance of these stations are evaluated on their effectiveness at intercepting potential human-trafficking victims given the amount of resources (e.g. staff, etc.) available.} \eb{These data enable comparison of border stations to determine which are efficient relative to other decision-making units (DMUs). Understanding the relative efficiency of border stations enables recommendations for best practices for improving operations. Operations Research methods applied in the context of anti-human trafficking are scarce, and this study, to the best of our knowledge, is the first to use DEA \citep{Dimas_2022}. In so doing, we illustrate how analytical methods can be applied to operational decision-making in the nonprofit environment.}

\eb{In what follows we overview how transit monitoring is used as a strategy to combat human trafficking and other public sector contexts in which DEA has successfully been applied. We then describe the data and DEA approach, homogeneity criteria, model inputs and outputs, and model results. We conclude with a list of  actionable recommendations as well as discuss the limitations and generalizability of the model. Additional model and data details can be found in the Appendices.}

\section{\eb{Transit Monitoring}}

\cite{Zimmerman_2011} depict trafficking as a series of event stages during which risks to an individual and intervention opportunities may arise: recruitment, travel-transit, exploitation, detention, integration, and re-trafficking. Although human trafficking, by definition, does not require movement of victims from one location to another, travel occurs in many human trafficking contexts and provides an opportunity for intervention.

\eb{The objective of transit monitoring is to identify and intercept those at risk of being trafficked in the travel-transit phase and prior to the exploitation phase (\citealt{Zimmerman_2011}, \citealt{Hudlow_2015}), so as to limit the level of trauma that any victims might experience. Trained personnel located along trafficking routes (such as transportation hubs and state border crossings) assess trafficking indicators and engage the potential victims in transit, involving government authorities where appropriate. \eg{Transit monitoring is more nuanced than the broader concept of network interdiction. Transit monitoring focuses on interrupting trafficking \emph{prior to} exploitation, whereas interdiction does not make this distinction.} Although it is possible to optimize strategic resource allocation decisions for network interdiction (including trafficking networks), the focus of this study is on the use of DEA to evaluate the effectiveness of LJI staff at conducting transit monitoring activities.}

\subsection{Human Trafficking and Transit Monitoring in Nepal}

Human trafficking in Nepal is a growing criminal industry \citep{USDS_2019}. Nepal is considered a “source country” of men, women, and children subjected to labor and sex trafficking \citep{USDS_2019}. The interaction of poverty, development, and relevant policies affect the vulnerability of its population to trafficking (\citealt{Joffres_2008}, \citealt{USDS_2019}). A large proportion of the population is estimated to be unemployed (42\%) or living below the poverty line (38\%). As of 2014, the most recent figures available, Nepal ranks 14\eg{2nd} out of 18\eg{9} countries on the Human Development Index \citep{UNDP_2020}; it has a low literacy rate (66\%), and more than 80\% of the country's inhabitants live in rural areas \citep{UNDP_2020}. Most trafficking victims in Nepal are female adults and children who not only suffer from gender inequality but are also particularly vulnerable due to economic insecurity, recent national disasters, and poverty (\citealt{Hennink_2004}, \citealt{Datta_2005}, \citealt{perry2013social}). Such factors generate high levels of migration to urban centers in Nepal, India, and the Middle East where trafficking victims are lured with promises of a better life, jobs, and false marriage proposals, or through the coercion of indebted families to sell their children (\citealt{Deane_2010}, \citealt{Kaufman_2011}, \citealt{perry2013social}). Migration is generally defined as the voluntary movement of persons within or across borders in search of a better livelihood. Trafficking and exploitation are associated with migration in two aspects: a person may willingly migrate for employment but find the work conditions to be exploitative; or may be deceived regarding the kind of work they will be doing (\citealt{Hennink_2004}, \citealt{Geiger_2010}).

In 1996, an anti-trafficking operation on Indian brothels rescued 200 female Nepali minors. \eb{This} occurrence \eb{spurred increased attention of} anti-human trafficking in Nepal. The \eb{Nepalese} government’s failure to help victims recover and rehabilitate motivated seven NGOs to take action \citep{Hudlow_2015}. Since 1996, these NGOs have played a significant role in combating human trafficking in Nepal, most notably by establishing transit monitoring and/or border monitoring stations along the largely unregulated Indian border. \eb{Although Nepal temporarily closed its international borders in 2020 due to the COVID-19 pandemic~\citep{Overseas_2020}, it is standard for the northern border with China to be tightly monitored, while the southern border with India is relatively open.} Although trade between Nepal and India is monitored through 22 checkpoint stations, Indian and Nepali individuals are permitted to cross the border at any point\eb{. In comparison, citizens} from other countries are only permitted to cross at 6 border checkpoint stations after obtaining entrance and exit visas \citep{Kansakar_2001}. The negligence and corruption of the border monitoring protocol have led to a prominent culture of transnational crime that includes human trafficking \citep{Uddin_2014}. In 2018, an estimated 171,000 Nepali individuals were victims of human trafficking or forced marriage \citep{GTI_2018}. The accuracy of the number of individuals trafficked in and out of Nepal is disputed, and the lack of precise and accurate data, an issue not only in Nepal but globally, is detrimental to measuring Nepal’s progress in combatting human trafficking \citep{Hudlow_2015,Dimas_2022}.

Love Justice International (LJI), formally known as Tiny Hands, is an NGO that combats human trafficking through transit monitoring methods. LJI focuses on preventing human trafficking \eg{prior to exploitation} by placing trained personnel at key transit points to identify people who exhibit \eg{risk indicators associated with currently being trafficked or being trafficked in the near future.} While LJI operates in 15 countries, the DEA model we present was developed for LJI operations in Nepal where transit monitoring practices were first launched in 2006\eb{,} along the Nepal-India border. Over the past \eb{14} years, LJI has \eb{grown to an operation of nearly} 30 monitoring station locations throughout Nepal\eb{.} Most stations are located on the Indian border, but some are located in interior major transportation hubs\eb{,} such as regional bus stations, and airports.

The DEA model enables LJI to identify both \eb{high} and \eb{low}-performing stations\eb{, where performance refers to} a station’s effectiveness of intercepting potential human-trafficking victims given the amount of \eg{available} resources (\eg{such as} staff). LJI is using the results of this study to develop best practices, evaluate and improve performance at inefficient stations, and more generally, use their limited anti-trafficking resources more efficiently.

\section{\eb{DEA Applications in The Public Sector}}
Few applications of analytical approaches towards anti-human trafficking operations currently exist (see, e.g., \citealt{Dubrawski_2015}, \citealt{Konrad_2017}, \citealt{Maass_2020}, \eb{\citealt{Dimas_2022}}), and to the best of our knowledge, there are no known DEA applications in anti-trafficking operations. DEA is useful to evaluate the relative efficiency of units within an organization that has different levels of inputs (e.g., resources, demand) that are used to produce outputs (e.g., productivity measures). \emph{Efficiency} refers to a unit’s ability to produce an expected amount of output given the amount of input resources. DEA models are employed in areas that range from the assessment of public organizations (such as healthcare systems, educational institutions, and governmental bodies) to private organizations (such as banks, restaurants, and service providers). \cite{ahn18} provide a recent review of the development and use of DEA models in the public sector. \eb{DEA has been applied to humanitarian efforts such as the work done in \cite{ald19} where the authors measured the effectiveness of humanitarian aid efforts across 106 different countries. Similarly,~\cite{Izadikhah_2019} considers the efficiency of humanitarian supply chains using a network DEA model.}

In a domain loosely related to transit monitoring, DEA has been used in policing to evaluate operational efficiencies. Aspects of policing work have some parallels with transit monitoring; namely, units of an organization or agency are seeking to \eb{identify} illicit behavior. For example, \cite{Sun_2002} evaluated the relative efficiency of Taiwanese police precincts and found that operational and process differences primarily depend on the resident population and location. \cite{Verma_2006}  measured the efficiency of state police units in India and found that a DEA model can generate targets of performance, identify inefficient departments, and determine adequate levels of operation and improvements in the units of criminal justice systems. Through DEA, \cite{Garc_a_S_nchez_2007} assessed the effectiveness of the Spanish police force to determine the most effective overall unit. Using an adjusted version of the DEA methodology, ‘Benefit-of-the-Doubt’, \cite{Verschelde_2012} advocate \eb{for} the use of a custom made operations research framework to evaluate citizen satisfaction with police effectiveness of community-oriented local police forces in Belgium.

DEA has also been used to measure the performance of nonprofits (\citealt{Vakkuri_2003}, \citealt{Kim_2018}). DEA emerged as an alternative method to measure the operational efficiency of nonprofits because it is capable of handling multiple performance metrics that are categorized as “inputs” and “outputs” to provide a single composite efficiency score for each nonprofit organization compared to other organizations that produce similar outputs. \cite{Ozbek_2015} evaluated the efficiency of five NGOs in Turkey dedicated to serving people suffering from wars, invasions, and natural disasters, supporting displaced refugees, as well as distributing other forms of aid without discrimination. \cite{Widiarto_2015} developed a DEA model to measure the efficiency of microfinance institutions and alleviate global poverty by extending financing to the poor for social and financial efficiency. In the second stage of this study, \cite{Widiarto_2015} compare the performance and factors that contribute to the efficiency of Islamic microfinance institutions. A challenge \eb{arising from} evaluating efficiency for nonprofits is that NGOs have a social objective, meaning they do not take part in competitive markets that use net income and rates of return as efficiency indicators \citep{Nunamaker_1985}. Therefore, many have used arbitrary measures of efficiency that involve developing weighting methods to quantify the outputs of a nonprofit \citep{Nunamaker_1985}.

\section{Data}
LJI \eg{staff who} conduct transit monitoring are trained in a multi-step process for potential victim identification: (1) visual identification of people traveling near transit stations to detect suspicious activity; (2) engagement with suspicious parties for heightened profiling to obtain more specific details \eg{--} should this rise to the level of suspected trafficking \eg{(either currently or in the near future)}, (3) interception and further questioning of potential victims or traffickers, culminating in a possible completion of an \eb{Interception} Report Form (IRF); (4) completion of a Victim \eb{Interview} Form (VIF) for each individual in the party; and (5) verification of the responses to help validate pertinent details. The verification processes usually require cultural knowledge, cross-checking data in LJI’s ``Fusion Center'' database (consisting of 9,000 individuals known or suspected of being involved in human trafficking), and following up with a third party by phone, such as parents, relatives, universities or employers \citep{Hudlow_2015}\eb{.}

After visual identification of suspicious activity, trained LJI staff intercept the individual(s) and administer an IRF. The IRF consists of a point-based system, called “red flags”, created by LJI to help guide staff in determining an occurrence of suspected trafficking. This point system assigns larger numeric values to questions more indicative of trafficking risk. For a more thorough explanation of this process, we refer the interested reader to \cite{Hudlow_2015}. 

When a party’s responses to questions on the IRF exceed the predetermined “red flag” threshold, trained LJI staff administer a VIF for each individual in the group \citep{Hudlow_2015}. LJI staff may then refer the potential victim and perpetrator, if present, to local law enforcement for follow\eb{-}up investigation and prosecution. If in LJI’s judgment, the “red flag” threshold is not exceeded and there is insufficient support for an occurrence of suspected trafficking, a VIF is not filled out and the party continues their travels without further intervention. \eb{LJI replaced the VIF with a similar form called the Case Information Form (CIF) in July 2018; for ease of prose, we hereafter jointly refer to these forms as VIF.}

If \emph{both} an IRF and a VIF have been completed for an individual, these paper forms are translated from Nepali to English and sent to data analysis staff at LJI. The data entry specialists enter IRF and VIF form data into a standard MS Excel workbook with rows representing observations (potential victims) and columns representing various features of interest (responses from the respective forms).

\begin{figure}
\FIGURE
{\includegraphics{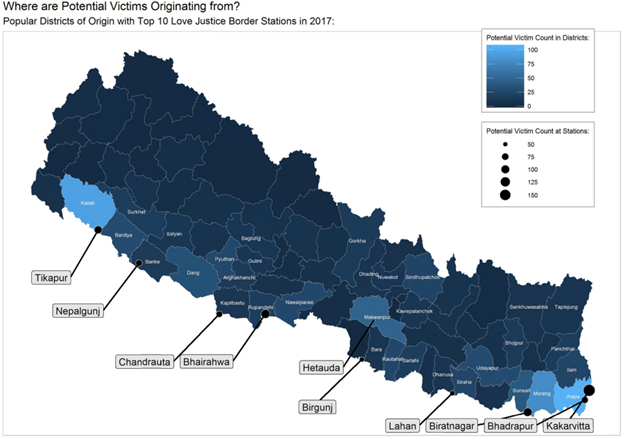}}
{\eb{The Top 10 Districts of Origin and LJI Border Stations, by Frequency of Victims, for 2017.}\label{fig1}}
{}
\end{figure}

We had access to over \eb{8,700} IRFs and \eb{nearly 4,500} VIFs ranging from late 2011 to 20\eb{21}. Collectively, the IRF and VIF data consist of over 350 data fields and capture both potential labor and sex trafficking transit activity. The data include detailed demographic information on the victims (age, gender, caste, health, marital status, history of abuse) and on the traffickers (age, gender, occupation, and how they met the victims). Rich information is also available regarding the trafficking supply networks, including victim origin and destination, recruitment methods, transit duration, methods of transportation, and safe houses. \eb{For example,} Figure \ref{fig1} \eb{depicts} the \eb{originating} districts \eb{of} potential trafficking victims \eb{who} were intercepted at LJI border stations in 2017 (for ease of \eb{prose}, only the 10 LJI border stations with the \eb{highest recorded} transit activity are displayed). 

Before data analysis and development of the DEA model, the data needed to be cleaned. This process involved linking the IRFs with their corresponding VIFs, organizing the data in a manner to reduce any redundancies, filling in and removing missing or incomplete data, and \eb{completing additional general data cleaning steps. Next, the data was filtered to find stations matching our homogeneity criteria and then each of the input and output features were aggregated on the quarter level as described in the methodology section below (see Table \ref{tab1}).} This process was conducted in the R scripting language.\\

\section{Methodology: Data Envelopment Analysis}
We employed DEA to quantify the relative efficiencies of \eb{deploying transit monitoring at various LJI stations in Nepal}. The purpose was to identify which LJI stations in Nepal are considered efficient and how the level of efficiency ranks among the stations. These results could then be used to identify potential methods of improving human trafficking transit monitoring tactics.
 
DEA is a mathematical programming method used to compute the relative efficiency of multiple decision-making units (DMUs), such as a hospital in a healthcare system, which have different inputs and outputs \citep{Charnes_1978}. The performance of \eb{each} DMU is evaluated by assessing \eb{its} ability to produce associated outputs by consuming its inputs\eb{, with respect to all other considered DMUs} \citep{Lim_2015}.

DEA models are commonly categorized by the shape of the identified best-practice frontiers. In the DEA literature, we use variable returns to scale (VRS) and constant returns to scale (CRS) to classify two basic DEA models. In the context of LJI transit monitoring, because the number of inputs does not result in a proportional change in outputs, a VRS model was most appropriate \eb{(see Appendix A for details)}. Moreover, our objective is to maximize the outputs achieved, and thus we employed an output-oriented VRS model \citep{ban84}; Appendix A details the algebraic formulation \eb{and further motivates our use of the VRS model}. We \eb{additionally} conducted a cross-efficiency analysis that allows for a quantified ranking of the \emph{relative} efficiencies of DMUs, which extends the findings of a standard DEA investigation that only identifies efficient or inefficient stations in a binary fashion \citep{Lim_2015}.

\begin{figure}
\FIGURE
{\includegraphics{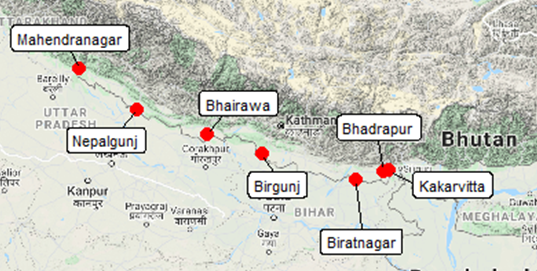}}
{\eb{The Seven LJI Border Crossing Stations Meeting the Qualifications for Inclusion in Our Analysis.}\label{fig2}}
{}
\end{figure}

It is generally required that all DMUs chosen must be homogenous under DEA. After all, as DEA aims to view all stations in their best relative light, it is only fair to compare DMUs that operate under relatively similar environments. In our context, the criteria we consider for homogeneity are \eb{1) the presence of a station manager; 2) the quality of data, and 3) flow through station locations.} Stations that shared similar ranges for these parameters were grouped. \eb{Each station considered for evaluation was required to have a station manager as these stations tend to be better managed and share similar operational standards.} The stations chosen were also required to \eb{have data (positive counts of IRF and VIF forms) for all consecutive quarters during our timeline to ensure reliability of the data.} \eb{Finally, we grouped stations with the same flow together for consideration. }\eb{ The station flow is the magnitude of total travelers through a station estimated from several factors, including the distance from the Indian border, number of airports, number of bus or train stations, number of major highways, and population density. Based on their scores in each category, a value of 1, 2, or 3 was assigned to the stations, with 1 and 3 representing the lowest and highest estimated flows, respectively.} For this analysis, \eb{we} considered \eb{only stations with a flow of 3, under the assumption} that \eb{larger} flow\eb{s} of people \eb{would lead to} an increased likelihood of intercepting trafficking activity. \eb{We validated our estimated flow ranking with our partners at LJI, which was well received.} Out of the eighteen stations for which we had data, \eb{seven} met these standards. \eb{Table \ref{tab1} summarizes the selection process for these stations.} The \eb{seven selected} stations \eb{for our analysis} are \eb{Mahendranagar}, Nepalgunj, \eb{Birgunj}, Kakarvitta, Biratnagar, Bhairawa, and Bhadrapur. Figure \ref{fig2} displays these stations geographically.\\



\begin{table}
\TABLE
{\eb{The DMU Selection Criteria and Process, Resulting in Seven Selected Stations for Further Analysis.}\label{tab1}}
{
\eb{\begin{tabular}{@{}ccc@{}}
\toprule
\rowcolor[HTML]{FFFFFF} 
\textbf{Step}                & \textbf{Selection Criteria}                                                                                                                  & \textbf{Total Stations Meeting Criteria} \\ \midrule
                             & \textbf{All Stations Operating Between 2011-2021}                                                                                            &                                          \\
\multirow{-2}{*}{\textbf{0}} & \down Number of stations observed in our dataset                                                                                                   & \multirow{-2}{*}{18}                     \\
                             &\up \textbf{Presence of a Station Manager}                                                                                                       &                                          \\
\multirow{-2}{*}{\textbf{1}} & \down Number of stations with a station manager on staff                                                                                           & \multirow{-2}{*}{12}                     \\
                             & \up \textbf{Quality of Data}                                                                                                                     &                                          \\
\multirow{-2}{*}{\textbf{2}} & \begin{tabular}[c]{@{}c@{}}Selected stations that operated consistently over time \\ \down(that is, have positive IRF and VIF form counts)\end{tabular} & \multirow{-2}{*}{11}                     \\
                             &\up \textbf{Flow Through Station Location}                                                                                                       &                                          \\
\multirow{-2}{*}{\textbf{3}} & \down Stations with a high level (3) of flow                                                                                                 & \multirow{-2}{*}{7}                      \\ \midrule
\multicolumn{3}{c}{\up\down \textbf{Final Dataset: 7 Stations * 13 Quarters = 91 Total DMUs}}                                                                                                                                   \\ \bottomrule
\end{tabular}}
}
{}
\end{table}

To ensure a sufficient number of \eb{DMUs} \citep{Podinovski_2007}, each of the \eb{seven} stations was expanded to \eb{thirteen} calendar quarters of data \eb{over} which we had complete information (from Q\eb{2} 2016 through Q\eb{2} 201\eb{9}). This created \eb{91} station-quarter DMUs\eb{, satisfying the empirical rule \eg{that suggests the number of DMUs be at least twice the number of inputs and outputs combined} \citep{Cook_2014}.} \eb{Because we use DEA as a tool for relative performance evaluation, rather than a production function estimate, the number of DMUs is much less critical for our study~\citep[see, e.g.,][]{Cook_2014}}. \eb{For ease of prose, we} refer to “station-quarters” as “stations” where unambiguous. \eb{While} our original data set contained information from late 2011 to 20\eb{21}, consistent, high-quality data for these \eb{seven} stations were only available for \eb{four} years; the first quarter in our analysis represents \eb{April} to \eb{June} of 2016, and the \eb{thirteenth} and last quarter in our analysis represents \eb{April} to \eb{June} of 201\eb{9}. \eb{We note that the COVID-19 pandemic impacted operations during 2020 and 2021 and therefore the data was incomplete and excluded from our analysis.}

DEA was applied to the \eb{91} LJI interception station-quarters to evaluate their performance in \eb{applying transit monitoring}. By assessing the efficiency of the station-quarters using the inputs consumed and outputs produced, LJI can begin to understand which station-quarters performed better at \eb{following procedures and} intercepting \eb{individuals at risk for} human trafficking. Such insights can then empower further root-cause analyses, which, in turn, provides LJI with a blueprint for \eb{improving} overall station performance.\\

\subsection{DEA Formulation to Evaluate LJI Transit Monitoring Stations}

Through discussions with LJI and \eb{availability of related} data, we identified the key inputs and outputs for our DEA modeling to evaluate the performance of human trafficking interception stations. \eb{We provide details of our final model below and refer the reader to Appendix B for an analysis of other considered inputs and outputs.}

{\bf Inputs.} We consider \eb{three} inputs for each station: \emph{number of staff}, \emph{test scores}, and \eb{\emph{hours worked by staff}}. The number of staff members represents the average number of LJI personnel employed by each station within a quarter, which ranged from \eb{2} to 12 individuals. Test scores represent the average staff performance on border knowledge exams, a mandatory exam each staff member must take before employment. The average quarterly test scores of the staff assigned to a station ranged from \eb{20} to 25. \eb{The average weekly hours worked by staff ranged from 36 to 50.}

{\bf Outputs.} We consider four outputs for each station: \emph{number of suspected trafficking occurrences} (count of IRFs), \emph{number of potential victims} (recorded on VIFs), \emph{IRF completeness}, and \emph{VIF completeness}.  \eb{T}here is one VIF for each individual recorded on the corresponding IRF form that exceeded the pre-determined trafficking risk threshold. The completeness of the IRFs and VIFs represents the thoroughness of staff filling out all of the minimum details necessary in each respective form.

In consultation with the LJI staff, IRF and VIF completeness were calculated as the number of forms that had filled out \eb{the required questions of each form.} If these questions were answered, a value of 1 was assigned to that question for that form, otherwise, a value of 0 was assigned. The completeness of each form \eb{is} then calculated by taking the percentage of the required questions \emph{answered} on a form out of the total questions required. \eb{As} the IRF asks questions regarding the collective group of people crossing the border together, a single IRF may contain information for multiple people. However, the VIF is completed individually for each person in the group. As a group of people crossing the border generates a single IRF and multiple VIFs, the VIF completeness measure associated with the encounter can be interpreted in multiple ways. For example, if three people are traveling together in a group the VIF completeness for the encounter could be measured by the minimum, maximum, average, or the sum of the completeness of the VIF forms for the three individual people. \eb{In consultations with} LJI stakeholders \eb{it was determined that} an encounter’s VIF completeness was \eb{most appropriately defined by} summing the VIF completeness of all individuals in the group.

\section{Findings and Recommendations}
\begin{figure}
\FIGURE
{\includegraphics[trim = 0 0 0 0, scale=1]{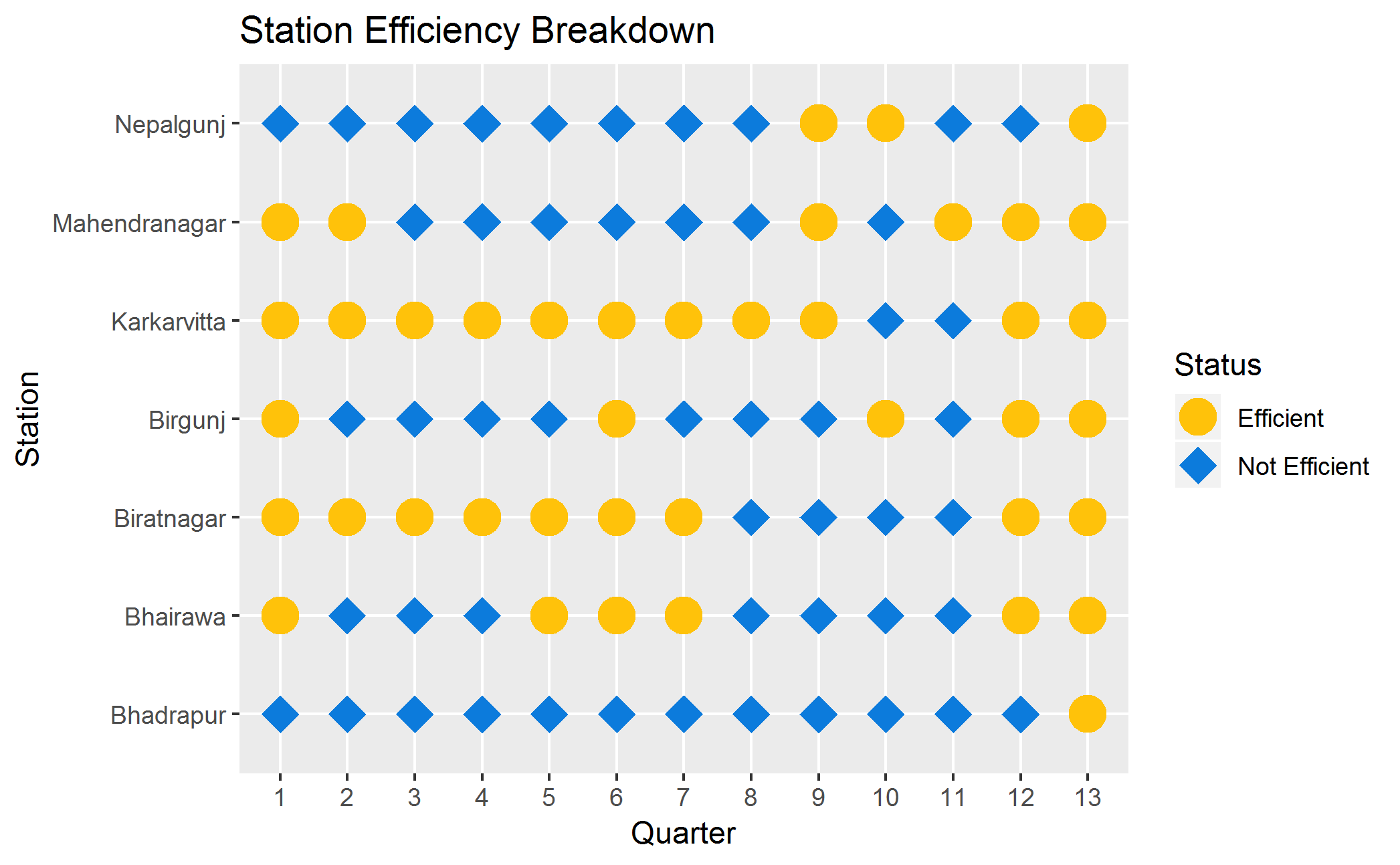}}
{\eb{Efficiency, by Station and Quarter, with Respect to the Best Practice Frontier.}\label{fig3}}
{}
\end{figure}

In LJI’s context, efficient stations are ones that have an adequate number of IRFs and VIFs filled out with a sufficient level of completeness \eb{for} each quarter given: (\eb{1}) the average number of staff employed at the station\eb{,} (\eb{2}) the staff’s border knowledge test scores, and (\eb{3}) \eb{the average weekly hours worked by staff.} A station is considered inefficient if it fails to produce adequate output from its characteristic input, indicated by a DEA model score of less than one. Figure \ref{fig3} illustrates the results of our \eb{initial} DEA investigations \eb{on station efficiencies}. \eb{Efficient stations are indicated by \eg{yellow} circles, while blue diamonds indicate inefficient stations. While no station of the seven was found to be} efficient every quarter, Kakarvitta and Biratnagar were efficient for \eb{a} majority of \eb{quarters}, with Kakarvitta having only \eb{two} out of \eb{thirteen} quarters identified as inefficient. \eb{The remaining \eg{five} stations} were inefficient for \eb{over} half of the quarters. \\

\eb{Of interest, is that all of the stations were efficient in quarter thirteen (Q2, 2019). While at first glance one might conclude that all stations have reached an efficient state, this is an incomplete assessment; variability across quarters reveals that efficiency can readily change from quarter to quarter.}

\begin{figure}
\FIGURE
{\includegraphics[trim = 0 0 0 0, scale=1.1]{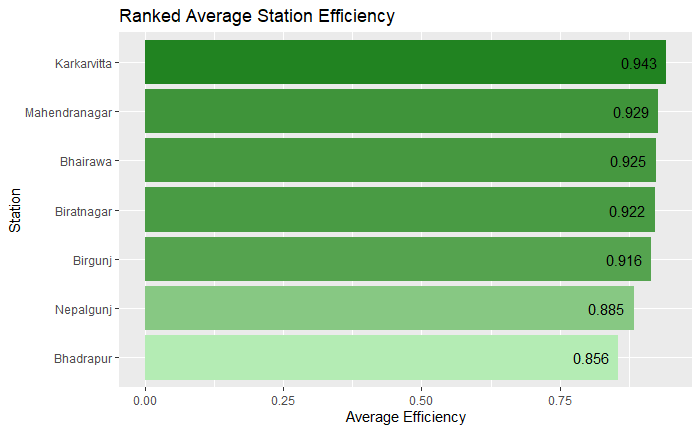}}
{\eb{Cross-Efficiency, by Station, Averaged over Quarters.}\label{fig4}}
{}
\end{figure}

\eb{Figure \ref{fig3} reveals DMUs on the best-practice frontier over 13 quarters and provides initial insights into a station’s own efficiency trajectory.} \eb{As station efficiency may vary over quarters, it would be incomplete to infer from Figure \ref{fig3} the best performing stations.} To understand the performance of \eb{individual stations with respect to others, we calculated} cross-efficiency values under variable RTS \citep{Lim_2015}. \eg{Cross-efficiency is a method to rank DMUs using weights that considers both peer and self-evaluation}  \eg{and} result\eg{s} i\eg{n} a ranking of stations based on their average cross-efficiency values over time, as compared to every other station. The average cross-efficiency based ranking for all \eb{seven} stations is shown in Figure \ref{fig4}. We note that while \eb{every} station had at least one quarter of inefficiency, the average cross-efficiency for each station is \eb{at least \eg{0.856}}.\\

\eb{Taken together, Figures \ref{fig3} and \ref{fig4} offer complementary views on station efficiency. Figure \ref{fig3} reveals quarter-by-quarter best-practice frontiers across the seven stations, while Figure \ref{fig4} reveals cross-efficiency scores for each station averaged across all quarters \citep{Lim_2015}. The resulting differences in Figures \ref{fig3} and \ref{fig4} with respect to rank and best-practice frontiers underscore the differences that can result from two different, yet related, comparison methodologies.}\\

\begin{figure}
\FIGURE
{\includegraphics[trim = 0 0 0 0, scale=0.95]{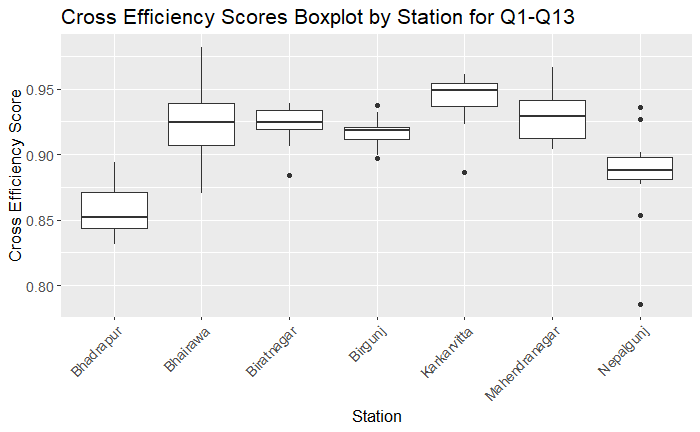}}
{\eb{Cross-Efficiency by Station, with Boxplots Depicting Distribution over 13 Quarters.}\label{fig5}}
{}
\end{figure}

\eb{Figure \ref{fig5} shows} quarter-to-quarter variation in cross efficiency scores. \eb{ Biratnagar, Birgunj, Karkarvitta}, and  \eg{Nepalgunj} \eb{have} the smallest \eb{interquartile ranges}, demonstrating low variance and consistency in performance. \eb{\eg{Bhadrapur}, Bhairawa, \eg{and} Mahendranagar} have larger interquartile ranges, indicating more performance variability between quarters. Figure \eb{\ref{fig5}} also substantiates that \eb{Nepalgunj} and Bhadrapur consistently produced the lowest cross efficiency scores.

\eb{Through careful interpretation, the results of a station's DEA performance can inform implementation toward the best-practice frontier, which can be seen as} the threshold in which a DMU is considered efficient based on its inputs and outputs. A station along the best-practice frontier indicates it is deemed efficient due to its ability to produce outputs with the given inputs. Based on our findings, we developed recommendations to increase the consistency of a station’s performance efficiency. These findings are based on numerical analysis from the DEA outputs, observations, and discussions with LJI. \eb{T}he numerical analysis refer\eb{s} to the DEA outputs \eb{that} describe the percentages that other efficient stations are contributing to the efficiency score of inefficient stations. The efficient output values are the dot product of the efficient stations’ output values and their percentage contribution of inefficiency. The DEA model results also disclosed the number of additional units of outputs that each station needs to produce to be considered efficient.

\begin{figure}
\FIGURE
{\includegraphics[trim = 0 0 0 0, scale=0.85]{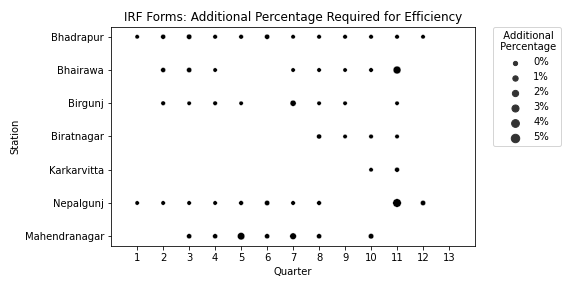}}
{\eb{The Additional Percentage of IRF Forms Needed for Efficiency Varies by Quarter and Station.}\label{fig6}}
{}
\end{figure}

\begin{figure}
\FIGURE
{\includegraphics[trim = 0 0 0 0, scale=0.85]{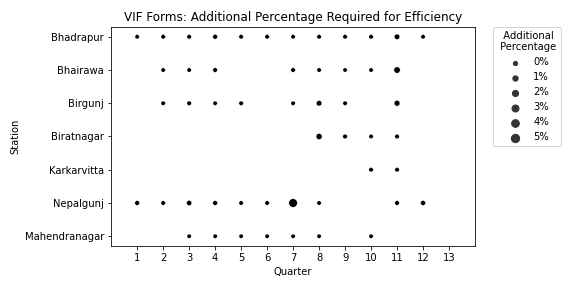}}
{\eb{The Additional Percentage of VIF Forms Needed for Efficiency Varies by Quarter and Station.}\label{fig7}}
{}
\end{figure}

\begin{figure}
\FIGURE
{\includegraphics[trim = 0 0 0 0, scale=0.85]{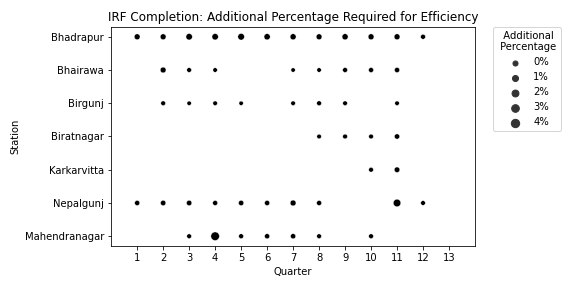}}
{\eb{The Additional Percentage of IRF Form Completion Needed for Efficiency Varies by Quarter and Station.}\label{fig8}}
{}
\end{figure}

\begin{figure}
\FIGURE
{\includegraphics[trim = 0 0 0 0, scale=0.85]{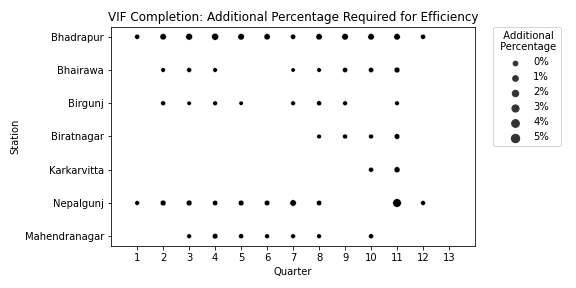}}
{\eb{The Additional Percentage of VIF Form Completion Needed for Efficiency Varies by Quarter and Station.}\label{fig9}}
{}
\end{figure}

\eb{Figures \ref{fig6} through \ref{fig9} show the results of this analysis for each output feature. The size of the circle depicts the percent of additional units required for a DMU to have been considered efficient. The absence of a circle indicates the outputs of a station were already operating on the best-practice frontier. For example in Figure \ref{fig6}, for Nepalgunj in quarter 11 to reach the efficiency frontier, an additional 2.5\% of IRF forms were required to be completed. These plots quickly reveal which stations require the largest increase in outputs.} Based on the\eb{se} values, we summarized observations from the numerical analysis for each station in Table \eb{\ref{tab2}}. \eb{We provide our recommendations by highlighting i) which stations require the most attention overall as well as ii) which outputs from each station need the most attention. Due to the limited amount of resources \eg{typically} available for NGOs such as LJI, these recommendations help provide insights on where to prioritize their resources and in which areas (outputs).  For each station (across all 13 quarters) we ranked which required the largest increase in outputs (Table \ref{tab2}, column 2). In addition, for each of these stations we ranked which outputs require the most attention (Table \ref{tab2}, columns 4 through 7) and summarized these results (Table \ref{tab2}, column 3). The ranking of the outputs range from 1 to 4, some outputs have equal rank and therefore may have the same value. For example, looking at the station Nepalgunj (Table \ref{tab2}, row 2) we see this station requires the most amount of improvement across all outputs. The output with the largest potential for improvement was the number of VIF forms, followed by both VIF completion and IRF forms, while IRF completion was the least important.}\\

\begin{table}
\TABLE
{\eb{Key Areas for Improvement, by Station, Yielding Recommendations for Inefficient Stations Toward Best-Practice Frontier.}\label{tab2}}
{
\eb{\begin{tabular}{@{}ccccccc@{}}
\toprule
\multicolumn{7}{c}{\textbf{Recommendations}}                                        \\ \midrule
\textbf{Station} & \multicolumn{1}{l}{\textbf{Overall Priority}} &\textbf{\begin{tabular}[c]{@{}c@{}}Key Areas \\ for Improvement\end{tabular}} & \textbf{\begin{tabular}[c]{@{}c@{}}VIF \\ Count\end{tabular}} & \textbf{\begin{tabular}[c]{@{}c@{}}VIF \\ Completeness\end{tabular}} & \textbf{\begin{tabular}[c]{@{}c@{}}IRF \\ Count\end{tabular}} & \textbf{\begin{tabular}[c]{@{}c@{}}IRF \\ Completeness\end{tabular}} \\ \midrule
\up\down Nepalgunj & \up\down 1  & \up\down VIF Forms  & \textbf{\up\down 1}  & \up\down 2 & \up\down 2 & \up\down 3 \\
\up\down Bhadrapur & \up\down 2 & \up\down VIF and IRF Completion & \up\down 2 & \textbf{\up\down 1} & \up\down 2 & \textbf{\up\down 1}\\
\up\down Mahendranagar & \up\down 3 & \up\down IRF Forms & \up\down 4 & \up\down 3 & \textbf{\up\down 1} & \up\down 2                             
\\\up\down Karkarvitta & \up\down 4 & \up\down VIF Completion & \up\down 3 & \textbf{\up\down 1} & \up\down 2  & \up\down 2\\\up\down Bhairawa & \up\down 5  & \up\down IRF Forms  & \up\down 2 & \up\down 4 & \textbf{\up\down 1}& \up\down 3 
\\\up\down Biratnagar & \up\down 6 & \up\down VIF Forms & \textbf{\up\down 1} & \up\down 2 & \up\down 4 & \up\down 3
\\\up\down Birgunj & \up\down 7& \up\down VIF Forms & \textbf{\up\down 1}
& \up\down 2 & \up\down 4 & \up\down 3 \\ \bottomrule
\end{tabular}
}
}
{}
\end{table}

To improve the performance of stations that lie outside of the best-practice frontier due to IRF \eb{and VIF completeness} (Table \eb{\ref{tab2}}, column\eb{s 5 and 7}), LJI should consider \eb{conducting} additional IRF \eb{and VIF} training and quality control checks on these stations. \eb{Recently, LJI has started to provide workers with a “cheat sheet” that emphasizes required questions on the IRF and VIF forms. The implementation of this new tool is expected to be an effective way to improve these outputs.} If the rate of trafficking were directly proportional to the rate of flow at all border stations, then \emph{mathematically} speaking, increasing IRF and VIF collection suggests that LJI would intercept more \eb{individuals at risk of} being trafficked-- that is, fewer \eb{potential} victims would be missed (Table \eb{\ref{tab2}}, column\eb{s 4 and 6}). As is so often the case in anti-human trafficking work, knowledge of the ground truth of actual trafficking cases is lacking. Hence, we instead used a proxy of overall flow, which represents not only trafficking occurrences, but also legal transit -- ordinary transnational travel, including those willingly traveling for gainful employment.

As such, care is required in interpreting recommendations \eg{before putting} them to practice. Rather than simply increas\eb{ing} IRF and VIF collection on the entire flow of travelers, it should be done \emph{conditional on actual occurrences of trafficking}. We recommend \eb{that} LJI conduct supplementary analysis to determine if any of these stations are expected to have a less-than-proportional number of potential trafficking victims passing through their station relative to the total number of travelers. Stations with a less-than proportional trafficking flow may not actually need to increase the number of IRF and VIF forms collected, as the low performance on this measure may be an artifact of fewer people being trafficked through this location.

That said, for stations where it is reasonable to assume that identification of additional potential trafficking victims will move them toward the best-practice frontier, we recommend LJI increase IRF and VIF collection on this targeted population of flow---actual trafficking occurrences---by improving strategies and tools for identification, as well as continuing education and training for staff. Specifically, LJI should focus on whether there are any cultural or geographical factors, such as recent changes in migration trends; a newly favored route or terrain to cross the border in nearby locations that LJI staff have been unable to monitor; or ways traffickers may be coaching victims on responses to LJI questions to evade detection that may be affecting the number of \eb{potential} victims identified at stations. While such additional analyses are outside the scope of the present study, they can provide useful supplementary information to LJI regarding where to focus their efforts for further analysis.

\section{Conclusion}
Human trafficking is a complex societal issue requiring a variety of approaches to help counter the social, health, and economic impacts associated with this crime. Transit monitoring has emerged as a promising strategy in reducing potential trafficking and exploitation activities, in which trained personnel are strategically \eb{located} along \eb{potential} trafficking routes to dynamically assess trafficking indicators and intercept probable victims \eg{prior to exploitation}. While transit monitoring is an effective strategy \eb{in preventing trafficking, transit monitoring} stations can vary in their performance, and there are opportunities to evaluate and improve stations by sharing best practice recommendations among stations. To this end, we used DEA to differentiate between efficient and inefficient stations over time. Our approach is innovative as a performance management methodology for nonprofits in the anti-trafficking sector as it allows decision-makers to evaluate their organizations and units with multiple inputs and outputs, benchmark units against comparable peers, and plan for the future based on a realistic allocation of resources. It is believed that this effort is the first application of DEA to evaluate performance in anti-human trafficking efforts.

In resource-constrained environments in which nonprofits often operate, analyses to support operational decisions offer an opportunity to improve operations. \eb{ While} such analyses have a long history of leading to more efficient operations in the for-profit sector\eb{, they appear less} frequently in the nonprofit sector. This \eb{study} presents the experiences from a \eb{collaboration} between operations research analysts and a nonprofit engaged in anti-human trafficking initiatives. \eb{We} discuss the use of DEA to examine the performance of stations engaging in transit monitoring in collaboration with Love Justice International \eg{(LJI)}, a nonprofit human trafficking organization engaged in transit-monitoring along the Nepal-India border. The approach and results were evaluated in several meetings \eb{with LJI} stakeholders\eb{, which} provided \eb{insights for better calibrating our model with reality.} \eb{This iterative process established our final model, and highlights} the significance of engaging all parties in determining \eb{model parameters (such as} inputs and outputs\eb{) to increase the} relevan\eb{ce of model outcomes (such as} performance evaluation\eb{)}. \eb{Our analysis} identified performance inefficiencies of individual transit monitoring stations. A repository of all R \eb{and python} code used to create the DEA model and data visualizations can be found publicly at \eb{[github link hidden to preserve blind review].}

\eb{This study is not without limitations, which may also present as opportunities for future investigations.} \eb{We} acknowledge that variables that were not included in our modeling and may affect its effectiveness. For example, \eb{other} transit-monitoring NGOs operate at \eb{certain} transit-monitoring stations, which could impact the inputs and outputs in our DEA model. We attempted to mitigate any confounding variables by basing our analysis on LJI’s consistent data format. Though there may be other factors to consider, we were able to develop a framework that can be expanded upon to help NGOs anti-trafficking efforts.

Nonprofits often have an abundance of routinely collected data. However, such data often lies idle, in locations that are inconvenient to access, in formats that are challenging to manipulate. Moreover, data quality issues may exist. For these reasons, there is a great opportunity for future research on improving the collection and management of data in the nonprofit and NGO sector. By doing so, NGO and nonprofit sector data can be better leveraged, providing an immense opportunity for improved operational analyses that are both accurate and meaningful.

While DEA produces meaningful recommendations, its mechanism and logic may \eb{prove challenging} to grasp for stakeholders lacking an analytical background. We used visual aids and verbal explanations over several NGO stakeholder meetings to illustrate our findings. DEA does have methodological drawbacks, such as difficulty in handling negative numbers, challenges with handling zeros and missing data in a straightforward manner, and being highly sensitive to outliers and bad data \citep{Medina_Borja_2007}. Some of these challenges were present in our study; as a result, some DMUs \eb{lacking positive counts of IRF and VIF forms during our timeframe needed to be omitted}, and only a subset of stations could be evaluated. Additionally, our decision to add quarters to increase the number of comparison DMUs made our analysis unique \eb{in enabling quarter-by-quarter comparisons of} stations with themselves. \eb{Although the use of quarter-by-quarter comparison increased the number of DMUs for analysis, the lower number of stations restricted the types of DEA analysis used in this work and is a limitation.} Another limitation of our analysis is that the most recent data we used was over \eb{a} year old \eb{due to impacts of the COVID-19 pandemic}. While this lag implies that more timely findings and actions are possible, the intent of the study \eb{was accomplished}: to build a framework upon which a nonprofit could measure performance and identify areas for improvement. The results of the DEA model confirmed anecdotal experiences of LJI operations. LJI is now empowered to take these results and this approach to identify the best-performing stations, understand what made other stations fall short, and implement best practices across all stations to increase effectiveness across the board. Additionally, the model can be used in the future with updated data to obtain timely results \eb{and inform continuous improvements in LJI’s transit monitoring efforts to combat the societal ill of human-trafficking.}\\

\ACKNOWLEDGMENT{%
We are grateful for the financial support from the National Science Foundation (Operations
Engineering grant CMMI- 1841893) including three students sponsored on REUs. We are indebted
to Samuel Longenbach, Biao Yin, Queting Wang, and Breanna Schad for their help in preparing
the data and conducting research for analysis. We are further indebted to Kirk Schweitzer \eb{ and Jon Hudlow }of Love Justice International for insightful discussions and comments on this study that have improved its implications.

}

%
%
%
 

\clearpage
\bibliographystyle{informs2014} 
\bibliography{ref1.bib} 


\begin{APPENDICES}
 \section{Appendix A. Data Envelopment Analysis}
 
 \eb{Data Envelopment Analysis (DEA) dates to a seminal paper by Charnes et al. (1978) and provides a mechanism to compare similar entities, known in DEA as decision-making units (DMUs), to identify those that are exhibiting (in)efficiency relative to other DMUs. All DMUs are evaluated on their efficiency by their consumption of the same inputs to produce the same outputs, in various quantities. The purpose of DEA is to allow for each DMU to express itself as efficient (with respect to comparable DMUs), through some combination of using its inputs to form outputs. Understanding the relative (in)efficiency of various DMUs enables best-practice recommendations to be developed for improving the operations of inefficient DMUs.

DEA models may consider returns to scale as constant (CRS) or variable (VRS). The former assumes that a change in the inputs results in a change of outputs in the same scale. The latter allows for non-constant (that is, increasing or decreasing) returns to scale. In the current study, as there is no clear indication that a change in the inputs would result in the output change in the same scale, we consider the VRS model.
}

Suppose there are $|J|$ decision making units (DMUs) that consume $|M|$ inputs to produce $|N|$ outputs. Specifically, DMU $ j \in J$ consumes $x_{m_j}$ units of input $m \in M$ to produce $y_{n_j}$ units of output $n \in N$. The output-oriented, VRS DEA model maximizes the efficiency of DMU $j_0 \in J$  (Banker, Charnes, and Cooper, 1984) by determining the optimal values of the input and output weights, $v_m$ and $u_n$, respectively, in the following linear program:

\begin{tabular}{ l l  }
\up\down \underline{Sets}\\
 \emph{\up\down $J$} & \up\down Set of decision making units  \\
  \emph\up\down {$M$} & \up\down Set of inputs  \\
  \emph{\up\down $N$} & \up\down Set of outputs  \\

\up\down \underline{Parameters}\\
 \emph{\up\down $x_{m_j}$} & \up\down Amount of input $m \in M$ from unit $j  \in J$  \\
  \emph{\up\down $y_{n_j}$} & \up\down Amount of output $n \in N $ from unit $j  \in J$  \\

\up\down \underline{Decision Variables}\\
 \emph{\up\down $w_m$} & \up\down Weight given to input $m  \in M$ \\
  \emph{\up\down $u_n$} & \up\down Weight given to output $n  \in N$  \\
  \emph{\up\down $v$} & \up\down Scale factor \\
\up\down \underline{Output-Oriented Formulation}
\end{tabular}



\eg{
\begin{subequations} \label{prob1}
	\begin{alignat}{2}
	\text{Minimize}	\quad & \sum_{m \in M} w_m x_{m j_0} - v \notag \\
	\text{Subject to:}
	\quad & \sum_{n \in N} u_n y_{n j_0} = 1, \notag \\
	\quad & \sum_{m \in M} w_m x_{mj} - \sum_{n \in N} u_n y_{n_j} - v \geq 0, \quad \ \forall \ j \in J \notag \\
  \quad & v \ \text{free}, \notag \\
  \quad & u_n, w_m \geq 0, \quad \ \forall \ n \in N, \ \forall \ m \in M. \notag 
	\end{alignat}	
\end{subequations}
}

\noindent \eb{For each DMU $j \in J$, denoted as the reference DMU $j_0$, the above linear program (LP) is solved with respect to $j_0$. The goal is to determine weights, or price multipliers, that optimize the efficiency objective while ensuring that the total weighted input contributions for DMUs must be at least as large as the total weighted output contributions (so that DMUs cannot receive more in outputs than is put in).
If there exists weights such that the weighted input contribution is equal to the weighted output contribution for the reference DMU (obtained through optimizing the LP), then this DMU is considered \emph{efficient}. If this is not possible, that is, if there do not exist weights that enable the DMU to get the value out that it puts in, then it is \emph{inefficient}. For every inefficient DMU, there exists some combination of efficient DMUs that outperforms the inefficient DMU.
}

 \section{\eb{Appendix B. Sensitivity Analysis of Input and Output Features}}

 \eg{
We ran additional experiments across varying inputs and outputs to check the robustness of our final selected model. 
} \eb{Table \ref{tab3} details the three inputs and five outputs considered throughout our modeling process. We considered multiple combinations prior to settling on our final model, which is model A in Table~\ref{tab4}. The most salient experiments and results are presented below in Tables~\ref{tab4} and~\ref{tab5}, respectively.}

\begin{table}
\TABLE
{\eb{Three inputs and five outputs were considered in the DEA model.}\label{tab3}}
{
\eb{
\begin{tabular}{@{}c|l|l@{}}
\toprule
\multicolumn{1}{l}{}& \multicolumn{1}{c}{Feature}& \multicolumn{1}{c}{\eg{Description}}   \\ \midrule
\multirow{3}{*}{Inputs}  & \up\down Number of Staff & \up\down The average number of staff working at a station per quarter. \\
\up\down & Staff Test Scores & \up\down The average test scores of staff working at a station per quarter.\\
&\up Hours Worked by Staff & \up The average weekly hours worked per staff in a quarter.     \\ \midrule
\multirow{6}{*}{Outputs} & \down Count of IRFs & \down The number of IRF forms collected per quarter.\\
& \up\down Count of VIFs & \up\down The number of VIF forms collected per quarter.\\
& \up\down IRF Completeness & \up\down The average completeness of IRF required questions per quarter.\\
& \up\down VIF Completeness & \up\down The average completeness of VIF required questions per quarter.\\
& \multirow{2}{*}{\begin{tabular}[c]{@{}l@{}}\up\down Percent of \eg{``Correct"}  \\ Instances of Trafficking\end{tabular}} & \begin{tabular}[c]{@{}l@{}}\up\down The percent of total VIF forms that are estimated to be a \\positive instance of trafficking.\\ \end{tabular} \\
& & \textit{\begin{tabular}[c]{@{}l@{}}It is important to note this feature is limited by the data available\\
and likely an overestimate.\end{tabular}}
\\ \bottomrule
\end{tabular}
}
}
{}
\end{table}

\eb{The first set of experiments, which we call \emph{varying input} experiments, are those we varied only the inputs, holding constant the four outputs used in our final model. \eg{We note that transit monitoring (application of the IRF and VIF forms) has two objectives: 1) to interact with a multitude of individuals (the count of forms) and 2) to sufficiently complete the forms (completeness measure) to help determine those who are at higher risk of being trafficked. Even so, we carried out experiments to exclude both VIF and IRF form counts and only include the form completeness measures. These results did have an effect on the cross-efficiency rankings of stations (see Table 5). Moreover, five DMUs that under our final proposed model are seen as “efficient”, were not “efficient” in this experiment. LJI’s operations are currently designed around the assumption that the more effective an interaction with individuals is (measuring in both counts and completeness), the better the chances of identifying a potential instance of trafficking, and therefore aiding in the fight against HT. Therefore, as we are only comparing stations with a similar amount of flow (people crossing the station), the count of forms is an important output measure for LJI’s transit monitoring efforts and included in our final model.} It should \eg{also} be noted that although both model A and E produce the same rankings, staff hours were deemed an important feature to include in this context and therefore the set of inputs considered hereafter are: \emph{number of staff}, \emph{staff test scores} and \emph{staff hours}.}

\eb{To test the sensitivity of this Model A with respect to the outputs, we ran \emph{varying output} experiments that varied only the outputs, holding the inputs constant. \eg{One such output considered was an approximate estimate of the proportion of ``correct instances" of trafficking. Given the limitations of available data that was collected, this value better represents an instance of an individual most likely to have been trafficked. This  approximation was calculated using two features in the data and while our estimate is far from perfect, this combined feature provides more information on those who are more likely to be a ``correct instance” of trafficking. Using this data we created an output feature that was the proportion of “correct instances” out of the total number of VIF forms filled out. The average percentage across all DMUs was just under 60\%. After using this feature for robustness checking, we determined that the cross-efficiency rankings were consistent whether this feature was included or excluded. As the results did not change, and this feature was an approximation, we retained the simpler model A (excluding this new output feature).}

Table~\ref{tab4} describes each model considered. We compared these different models on the cross-efficiency scores of our final output-oriented VRS model (base model). The results are presented in Table~\ref{tab5}; the parenthetical number is the ranking for the highest cross-efficiency, with one indicating the highest (most) station efficiency, and seven indicating the lowest (least). Bolded values indicate instances where a station’s ranking was impacted by model variation. These experiments resulted in model A being the best suited model for LJI, and thus the final model presented in our study. 
}

\begin{table}
\TABLE
{\eb{Twelve experiments were considered by varying the input and outputs.}\label{tab4}}
{
\eb{
\begin{tabular}{@{}lcccc|ccccc@{}}
\toprule & \multicolumn{1}{l}{} & \multicolumn{3}{c|}{\textbf{Input Values}} & \multicolumn{5}{c}{\textbf{Output Values}}\\ \midrule
\multirow{-2}{*}{} & \multicolumn{1}{c|}{\textbf{Model}} & \multicolumn{1}{c|}{\textbf{\begin{tabular}[c]{@{}c@{}}Number \\ of Staff\end{tabular}}} & \multicolumn{1}{c|}{\textbf{\begin{tabular}[c]{@{}c@{}}Staff Test\\ Scores\end{tabular}}} & \multicolumn{1}{c|}{\textbf{\begin{tabular}[c]{@{}c@{}}Staff\\ Hours\end{tabular}}} & \multicolumn{1}{c|}{\textbf{\begin{tabular}[c]{@{}c@{}}IRF\\ Forms\end{tabular}}} & \multicolumn{1}{c|}{\textbf{\begin{tabular}[c]{@{}c@{}}VIF\\ Forms\end{tabular}}} & \multicolumn{1}{c|}{\textbf{\begin{tabular}[c]{@{}c@{}}IRF\\ Completion\end{tabular}}} & \multicolumn{1}{c|}{\textbf{\begin{tabular}[c]{@{}c@{}}VIF \\ Completion\end{tabular}}} & \multicolumn{1}{c|}{\textbf{\begin{tabular}[c]{@{}c@{}}\eg{``Correct"}\\ Instance of\\ Trafficking\end{tabular}}} \\ \midrule
& \cellcolor[HTML]{C0C0C0}A & \cellcolor[HTML]{C0C0C0}\textbf{$\bullet$} & \cellcolor[HTML]{C0C0C0}\textbf{$\bullet$} & \cellcolor[HTML]{C0C0C0}\textbf{$\bullet$}& \cellcolor[HTML]{C0C0C0}\textbf{$\bullet$}& \cellcolor[HTML]{C0C0C0}\textbf{$\bullet$}& \cellcolor[HTML]{C0C0C0}\textbf{$\bullet$}& \cellcolor[HTML]{C0C0C0}\textbf{$\bullet$}& \cellcolor[HTML]{C0C0C0}\textbf{}\\[0em] 
\midrule
& B & $\bullet$ & & & $\bullet$ & $\bullet$  & $\bullet$  & $\bullet$ & \\[0.5em] 
& C & & $\bullet$  & & $\bullet$ & $\bullet$ & $\bullet$ & $\bullet$ & \\[0.5em] 
& D & & & $\bullet$ & $\bullet$ & $\bullet$ & $\bullet$ & $\bullet$ & \\[0.5em] 
& E & $\bullet$ & $\bullet$ & & $\bullet$ & $\bullet$ & $\bullet$ & $\bullet$ &\\[0.5em] 
& F & $\bullet$ & & $\bullet$ & $\bullet$ & $\bullet$ & $\bullet$ & $\bullet$ & \\[0.5em] 
\multirow{-7}{*}{\rotatebox[origin=c]{90}{Varying Inputs}} 
& G & & $\bullet$ & $\bullet$ & $\bullet$ & $\bullet$ & $\bullet$ & $\bullet$ & \\ [0.5em] \midrule
& H & $\bullet$ & $\bullet$ & $\bullet$ & & & $\bullet$ & $\bullet$ &\\[0.5em] 
& I & $\bullet$ & $\bullet$ & $\bullet$ & & & $\bullet$ & $\bullet$ & $\bullet$ \\[0.5em] 
& J & $\bullet$ & $\bullet$ & $\bullet$ & $\bullet$ & $\bullet$ & $\bullet$ & $\bullet$ & $\bullet$\\[0.5em] 
\multirow{-5}{*}{\rotatebox[origin=c]{90}{Varying Outputs}} 
& K & $\bullet$ & $\bullet$ & $\bullet$ & $\bullet$ & $\bullet$ & & & $\bullet$ \\[0.5em] 
& L & $\bullet$ & $\bullet$ & $\bullet$ & $\bullet$ & $\bullet$ & & & \\[0.5em]                         \bottomrule
\end{tabular}}
}
{}
\end{table}

\begin{table}
\TABLE
{\eb{The Cross-Efficiency of the Twelve Experiments, Revealing which Stations have the Highest and Lowest Cross-Efficiency.}\label{tab5}}
{
\eb{
\begin{tabular}{@{}ll|ccccccc@{}}
\multicolumn{2}{l}{} & \multicolumn{7}{c}{\textbf{Station}}\\ \midrule
\multicolumn{2}{l|}{\multirow{-1}{*}{\textbf{Model}}} 
& \multicolumn{1}{l|}{\textbf{Nepalgunj}}& \multicolumn{1}{l|}{\textbf{Mahendranagar}}  & \multicolumn{1}{l|}{\textbf{Karkarvitta}} & \multicolumn{1}{l|}{\textbf{Birgunj}}   & \multicolumn{1}{l|}{\textbf{Biratnagar}}& \multicolumn{1} {l|}{\textbf{Bhairawa}}    & \multicolumn{1}{l}{\textbf{Bhadrapur}}\\ \midrule
& \cellcolor[HTML]{C0C0C0}A  & \cellcolor[HTML]{C0C0C0}\textbf{\begin{tabular}[c]{@{}c@{}}0.885\\ (6)\end{tabular}} & \cellcolor[HTML]{C0C0C0}\textbf{\begin{tabular}[c]{@{}c@{}}0.929\\ (2)\end{tabular}} & \cellcolor[HTML]{C0C0C0}\textbf{\begin{tabular}[c]{@{}c@{}}0.943\\ (1)\end{tabular}} & \cellcolor[HTML]{C0C0C0}\textbf{\begin{tabular}[c]{@{}c@{}}0.916\\ (5)\end{tabular}} & \cellcolor[HTML]{C0C0C0}\textbf{\begin{tabular}[c]{@{}c@{}}0.922\\ (4)\end{tabular}} & \cellcolor[HTML]{C0C0C0}\textbf{\begin{tabular}[c]{@{}c@{}}0.925\\ (3)\end{tabular}} & \cellcolor[HTML]{C0C0C0}\textbf{\begin{tabular}[c]{@{}c@{}}0.856\\ (7)\end{tabular}}
\\ \midrule
& \textbf{B}& \textbf{\begin{tabular}[c]{@{}c@{}}0.927\\ (5)\end{tabular}}          & \textbf{\begin{tabular}[c]{@{}c@{}}0.922\\ (6)\end{tabular}}                      & \textbf{\begin{tabular}[c]{@{}c@{}}0.949\\ (3)\end{tabular}}                      & \textbf{\begin{tabular}[c]{@{}c@{}}0.949\\ (2)\end{tabular}}                      & \textbf{\begin{tabular}[c]{@{}c@{}}0.950\\ (1)\end{tabular}}                      & \textbf{\begin{tabular}[c]{@{}c@{}}0.944\\ (4)\end{tabular}}                      & \begin{tabular}[c]{@{}c@{}}0.901\\ (7)\end{tabular}\\
& \textbf{C} & \begin{tabular}[c]{@{}c@{}}0.891\\ (6)\end{tabular}                  & \textbf{\begin{tabular}[c]{@{}c@{}}0.932\\ (4)\end{tabular}}                      & \begin{tabular}[c]{@{}c@{}}0.953\\ (1)\end{tabular}                               & \begin{tabular}[c]{@{}c@{}}0.924\\ (5)\end{tabular}                               & \textbf{\begin{tabular}[c]{@{}c@{}}0.935\\ (2)\end{tabular}}                      & \begin{tabular}[c]{@{}c@{}}0.934\\ (3)\end{tabular}                               & \begin{tabular}[c]{@{}c@{}}0.868\\ (7)\end{tabular}\\
& \textbf{D} & \begin{tabular}[c]{@{}c@{}}0.913\\ (6)\end{tabular}                  & \textbf{\begin{tabular}[c]{@{}c@{}}0.914\\ (5)\end{tabular}}                      & \begin{tabular}[c]{@{}c@{}}0.954\\ (1)\end{tabular}                               & \textbf{\begin{tabular}[c]{@{}c@{}}0.919\\ (4)\end{tabular}}                      & \textbf{\begin{tabular}[c]{@{}c@{}}0.949\\ (2)\end{tabular}}                      & \begin{tabular}[c]{@{}c@{}}0.939\\ (3)\end{tabular}                               & \begin{tabular}[c]{@{}c@{}}0.883\\ (7)\end{tabular} \\
& \textbf{E} & \begin{tabular}[c]{@{}c@{}}0.893\\ (6)\end{tabular}                  & \begin{tabular}[c]{@{}c@{}}0.933\\ (2)\end{tabular}                               & \begin{tabular}[c]{@{}c@{}}0.945\\ (1)\end{tabular}                               & \begin{tabular}[c]{@{}c@{}}0.929\\ (5)\end{tabular}                               & \begin{tabular}[c]{@{}c@{}}0.929\\ (4)\end{tabular}                               & \begin{tabular}[c]{@{}c@{}}0.931\\ (3)\end{tabular}                               & \begin{tabular}[c]{@{}c@{}}0.867\\ (7)\end{tabular}\\
\multirow{-7}{*}{\rotatebox[origin=c]{90}{Varying Inputs}}  
& \textbf{F}& \begin{tabular}[c]{@{}c@{}}0.920\\ (6)\end{tabular}                   & \textbf{\begin{tabular}[c]{@{}c@{}}0.924\\ (5)\end{tabular}}                      & \begin{tabular}[c]{@{}c@{}}0.948\\ (1)\end{tabular}                               & \textbf{\begin{tabular}[c]{@{}c@{}}0.938\\ (4)\end{tabular}}                      & \textbf{\begin{tabular}[c]{@{}c@{}}0.946\\ (2)\end{tabular}}                      & \begin{tabular}[c]{@{}c@{}}0.939\\ (3)\end{tabular}                               & \begin{tabular}[c]{@{}c@{}}0.891\\ (7)\end{tabular}\\
& \textbf{G} & \begin{tabular}[c]{@{}c@{}}0.881\\ (6)\end{tabular}& \textbf{\begin{tabular}[c]{@{}c@{}}0.926\\ (4)\end{tabular}}      & \begin{tabular}[c]{@{}c@{}}0.949\\ (1)\end{tabular}                               & \begin{tabular}[c]{@{}c@{}}0.908\\ (5)\end{tabular}                               & \textbf{\begin{tabular}[c]{@{}c@{}}0.927\\ (3)\end{tabular}}                      & \textbf{\begin{tabular}[c]{@{}c@{}}0.928\\ (2)\end{tabular}}                      & \begin{tabular}[c]{@{}c@{}}0.854\\ (7)\end{tabular}\\ \midrule
& \textbf{H} & \begin{tabular}[c]{@{}c@{}}0.917\\ (6)\end{tabular}                  & \textbf{\begin{tabular}[c]{@{}c@{}}0.947\\ (3)\end{tabular}}                      & \begin{tabular}[c]{@{}c@{}}0.956\\ (1)\end{tabular}                               & \textbf{\begin{tabular}[c]{@{}c@{}}0.947\\ (4)\end{tabular}}                      & \textbf{\begin{tabular}[c]{@{}c@{}}0.952\\ (2)\end{tabular}}                      & \textbf{\begin{tabular}[c]{@{}c@{}}0.944\\ (5)\end{tabular}}                      & \begin{tabular}[c]{@{}c@{}}0.891\\ (7)\end{tabular}\\
& \textbf{I}& \begin{tabular}[c]{@{}c@{}}0.891\\ (6)\end{tabular}                   & \begin{tabular}[c]{@{}c@{}}0.936\\ (2)\end{tabular}                               & \begin{tabular}[c]{@{}c@{}}0.937\\ (1)\end{tabular}                               & \textbf{\begin{tabular}[c]{@{}c@{}}0.923\\ (4)\end{tabular}}                      & \textbf{\begin{tabular}[c]{@{}c@{}}0.927\\ (3)\end{tabular}}                      & \textbf{\begin{tabular}[c]{@{}c@{}}0.921\\ (5)\end{tabular}}                      & \begin{tabular}[c]{@{}c@{}}0.861\\ (7)\end{tabular}\\

& \textbf{J}& \begin{tabular}[c]{@{}c@{}}0.853\\ (6)\end{tabular}                   & \begin{tabular}[c]{@{}c@{}}0.916\\ (2)\end{tabular}                               & \begin{tabular}[c]{@{}c@{}}0.922\\ (1)\end{tabular}                               & \begin{tabular}[c]{@{}c@{}}0.889\\ (5)\end{tabular}                               & \begin{tabular}[c]{@{}c@{}}0.892\\ (4)\end{tabular}                               & \begin{tabular}[c]{@{}c@{}}0.900\\ (3)\end{tabular}                               & \begin{tabular}[c]{@{}c@{}}0.823\\ (7)\end{tabular}\\
\multirow{-5}{*}{\rotatebox[origin=c]{90}{Varying Outputs}} 
& \textbf{K}& \begin{tabular}[c]{@{}c@{}}0.240\\ (6)\end{tabular}                   & \begin{tabular}[c]{@{}c@{}}0.433\\ (2)\end{tabular}                               & \begin{tabular}[c]{@{}c@{}}0.528\\ (1)\end{tabular}                               & \begin{tabular}[c]{@{}c@{}}0.301\\ (5)\end{tabular}                               & \begin{tabular}[c]{@{}c@{}}0.304\\ (4)\end{tabular}                               & \begin{tabular}[c]{@{}c@{}}0.431\\ (3)\end{tabular}                               & \begin{tabular}[c]{@{}c@{}}0.823\\ (7)\end{tabular}\\

& \textbf{L}& \begin{tabular}[c]{@{}c@{}}0.240\\ (6)\end{tabular} & \textbf{\begin{tabular}[c]{@{}c@{}}0.418\\ (3)\end{tabular}}     & \begin{tabular}[c]{@{}c@{}}0.516\\ (1)\end{tabular}                               & \begin{tabular}[c]{@{}c@{}}0.271\\ (5)\end{tabular}                               & \begin{tabular}[c]{@{}c@{}}0.313\\ (4)\end{tabular}                               & \textbf{\begin{tabular}[c]{@{}c@{}}0.442\\ (2)\end{tabular}}                      & \begin{tabular}[c]{@{}c@{}}0.218\\ (7)\end{tabular} \\ 
 \bottomrule
\end{tabular}}}
{}
\end{table}

 \end{APPENDICES}

\end{document}